\documentclass[useAMS,usenatbib]{mn2e}
\usepackage{graphicx}
\usepackage{color}
\usepackage{amssymb}
\usepackage{amsmath}
\usepackage{upgreek}
\usepackage{subfigure}
\usepackage{rotating}
\usepackage{pdflscape}
\usepackage{bm} % This makes letters bold and italic as needed in mathmode
\usepackage{mathtools}

\usepackage{comment}
\includecomment{details}
\graphicspath{{images/}}

\title[{Stellar cosmic rays in protoplanetary disks}]{{Stellar cosmic rays as an important source of ionisation in protoplanetary disks: a disk mass dependent process}}
\author[Rodgers-Lee, Taylor, Downes, Ray]{D. Rodgers-Lee$^{1,2}$\thanks{E-mail:
drodgers@tcd.ie}, A. M. Taylor$^{3}$, T. P. Downes$^{4}$, T. P. Ray$^{5}$ \\
$^{1}$ School of Physics, Trinity College Dublin, University of Dublin, College Green, Dublin 2, Co. Dublin, D02 PN40, Ireland \\
$^{2}$ Centre for Astrophysics Research, School of Physics, Astronomy and Mathematics, University of Hertfordshire, College Lane,\\ Hatfield AL10 9AB, UK \\
$^{3}$ DESY, D-15738 Zeuthen, Germany\\
$^{4}$ Centre for Astrophysics \& Relativity, School of Mathematical Sciences, Dublin City University, Glasnevin, D09 W6Y4, Ireland\\
$^{5}$ School of Cosmic Physics, Dublin Institute for Advanced Studies, 31 Fitzwilliam Place, Dublin 2,  D02 XF86, Ireland }
\begin{document}
\date{Accepted xxxx xxxxxx xx. Received xxxx xxxxxx xx; in original form xxxx xxx xx}
\pagerange{\pageref{firstpage}--\pageref{lastpage}} \pubyear{xxxx}
\maketitle

\label{firstpage}

\begin{abstract}

We assess the ionising effect of low energy protostellar cosmic rays in protoplanetary disks around a young solar mass star for a wide range of disk parameters. We assume a source of low energy cosmic rays located close to the young star which travel diffusively through the protoplanetary disk. We use observationally inferred values from nearby star-forming regions for the total disk mass and the radial density profile. We investigate the influence of varying the disk mass within the observed scatter for a solar mass star. 

We find that for a large range of disk masses and density profiles that protoplanetary disks are ``optically thin'' to low energy ($\sim$3\,GeV) cosmic rays. At $R\sim10$\,au, for all of the disks that we consider ($M_\mathrm{disk}=6.0\times10^{-4} - 2.4\times 10^{-2}M_\odot$), the ionisation rate due to low energy stellar cosmic rays is larger than that expected from unmodulated galactic cosmic rays. This is in contrast to our previous results which assumed a much denser disk which may be appropriate for a more embedded source. 

At $R\sim70$\,au, the ionisation rate due to stellar cosmic rays dominates in $\sim$50\% of the disks. These are the less massive disks with less steep density profiles. At this radius there is at least an order of magnitude difference in the ionisation rate between the least and most massive disk that we consider. 

Our results indicate, for a wide range of disk masses, that low energy stellar cosmic rays provide an important source of ionisation at the disk midplane at large radii ($\sim$70\,au).

\end{abstract}


\section{Introduction}
\label{sec:intro}

Young low-mass stars are surrounded by disks of dust and gas, so-called protoplanetary disks. These disks evolve on a timescale of a few Myrs and are known to accrete onto the central star. It seems likely that this accretion is facilitated via magnetic fields, either the magneto-rotational instability \citep[MRI,][]{balbus_1991} and/or magneto-centrifugally launched winds \citep{blandford_1982}. Both of these magnetically-driven physical processes require a certain level of ionisation for the neutral material to effectively couple to the magnetic field via ionised charge carriers. 

Indeed, the nature of this coupling between the field and the bulk fluid is critically dependent on the ionisation fraction, with Ohmic resistivity, the Hall effect and ambipolar diffusion all thought to be important in different parts of the disk \citep{wardle_2007}.  Much work has been published  on the impact of these effects on global disk dynamics: \citet{okeeffe_2014} using multi-fluid magneto-hydrodynamic (MHD) simulations and \citet{lesur_2014} using single-fluid MHD simulations, were the first to explore the dynamics of protoplanetary disks in the presence of all three of these effects, while others \citep[e.g.][]{bai_2014, gressel_2015, rodgers-2016, bethune_2017} followed up with explorations of disk dynamics and wind launching with a variety of distributions for the non-ideal MHD effects in the disk.  Since the size of the three relevant non-ideal MHD effects is so dependent on the ionisation fraction, all of these works suffer from the large existing uncertainty in the ionisation rate. Thus, determining the sources of ionisation in protoplanetary disks is of great importance.

While stellar X-rays \citep{ercolano_2013} and radioactivity \citep{umebayashi_2009} offer a certain amount of ionisation they fail to sufficiently ionise deep into the disk, in the region therefore commonly known as the dead-zone \citep{gammie_1996}. Due to the possibility that Galactic cosmic rays are suppressed by the young star's heliosphere \citep[as noted by][]{gammie_1996,glassgold_1997,cleeves-2013} some interest has been generated concerning the ionising effect of low energy cosmic rays ($\sim$GeV energies) produced by the young star itself \citep{turner_2009,rab_2017,rodgers-lee_2017,fraschetti_2018}. 

Stellar cosmic rays may be able to ionise the protoplanetary disk sufficiently to decrease the extent of the dead-zone. Therefore, the MRI may operate in larger regions of the disk than previously thought. This additional source of ionisation might also help to launch magneto-centrifugally winds at lower heights above the disk midplane.

The ionisation rate in protoplanetary disks is a difficult quantity to observationally constrain and yet remains the clearest indication of the presence of low energy cosmic rays, especially if other sources of ionisation in the system can simultaneously be accounted for. \citet{cleeves-2015} present a constraint on the total ionisation rate for TW Hya ($\zeta\lesssim 10^{-19}\mathrm{s^{-1}}$) using chemical modelling of $\mathrm{HCO}^+$ and $\mathrm{N_2H^+}$ observations. On the other hand, \citet{schwarz_2018} find that it is difficult to model current CO observations which find low CO abundances without cosmic rays (Galactic or otherwise) or UV photons. It is possible that low energy stellar cosmic rays could play a role in explaining these CO observations. Further away from the protoplanetary disk, \citet{ainsworth_2014} have found indications of non-thermal processes occuring in the jet bow shocks from young low-mass stars.

On the theoretical side, there has been much interest in the acceleration of stellar cosmic rays and their subsequent transport. \citet{padovani_2015} and \citet{padovani_2016} investigated possible acceleration sites in the vicinity of young stellar objects. \citet{padovani_2018} investigated the energy loss processes of galactic cosmic rays in very dense media which lead to higher ionisation rates as a function of grammage than expected from previous calculations \citep[such as][]{umebayashi_1981} which is also relevant for stellar cosmic rays. On a more global scale, \citet{gaches_2018} estimated the ionisation rate in star-forming molecular clouds due to energetic particles accelerated in the accretion columns of young stellar objects.

Regarding the transport of stellar cosmic rays, \citet{rodgers-lee_2017}, \citet{rab_2017} and \citet{fraschetti_2018} all found that low energy stellar cosmic rays injected near the star were unable to travel far into the protoplanetary disk. In the first case the cosmic rays lost their energy due to the high density in the inner regions of the disk near the midplane despite the assumption of diffusive transport. In \citet{rab_2017} rectilinear transport was assumed for mainly MeV cosmic rays. Thus, the combination of very low energy cosmic rays (which have high energy loss rates) with the geometric dilution factor of $1/r^2$ caused the cosmic rays to be attenuated significantly. Last, \citet{fraschetti_2018} using test particle simulations found that stellar cosmic rays were magnetically confined due to the stellar magnetosphere. 

It is important to note that these simulations all use a different value for the mass of the central star. \citet{rodgers-lee_2017}, \citet{rab_2017} and \citet{fraschetti_2018} use $M_*= 1, \,0.7$ and $ 1.35(\pm0.15) \, M_\odot$ for the young star, respectively. For \citet{rodgers-lee_2017} and \citet{rab_2017}, who implement a density profile for the disk, the scale height of the disk is dependent on the mass of the central star which will affect the absorption of the stellar cosmic rays.

As noted above each of these models assumed a different mass for the central star. Other differences also exist concerning the mass and density structure of the disk itself.
The density profile used in \citet{rodgers-lee_2017} is a radial power law profile (their Eq.\,7) with an exponential drop-off in the vertical direction. \citet{rab_2017} use a similar expression for the density profile but with a tapering-off radius at 100\,au and a disk mass of $0.01\,M_\odot$. On the other hand, \citet{fraschetti_2018} assume a more simplified disk model which lies in the $xy$-plane and has no vertical or radial density structure.

The aim of this paper is to build on our previous work assuming diffusive transport and to investigate different disk masses and density profiles based on physical parameters from current observations. We then examine the range of parameters which can increase the ionisation rate deep in the disk. The model and parameters used are described in Section\,\ref{sec:form}. In Section\,\ref{sec:results} we present our results and discuss them in the context of our previous findings, as well as in comparison to the literature, in Section\,\ref{sec:discussion}. Finally, we outline our conclusions in Section\,\ref{sec:conclusions}.

\section{Formulation}
\label{sec:form}
We use the same model as described in \citet{rodgers-lee_2017}. Namely, we assume diffusive transport for the low energy cosmic rays and solve the associated 2D transport equation
\begin{equation}
\frac{\partial n_\mathrm{CR}}{\partial t} = \bm{\nabla}\cdot(D\bm{\nabla} n_\mathrm{CR}) - \frac{n_\mathrm{CR}}{\tau} + Q
\label{eq:ncr}
\end{equation}
\noindent where $n_\mathrm{CR}$ is the number density of cosmic rays, $D(R,z)$ is the spatial diffusion coefficient, $1/\tau(R,z)$ is the cosmic ray energy loss rate \citep[given by Eq.\,6 in][ and is related to the mass density of the protoplanetary disk]{rodgers-lee_2017} and $Q$ is the injection rate of low energy cosmic rays. $Q$ is linked to the luminosity of low energy cosmic rays that we assume which is given in Section\,\ref{subsubsec:lcr}. The assumed diffusion coefficient is also given in Section\,\ref{subsubsec:lcr}. It is important to note that we do not consider an advective term here since we investigated this previously and found it to have little effect. Eq.\,\ref{eq:ncr} is solved using cylindrical coordinates assuming axial symmetry. Further details regarding the code are given in \citet{rodgers-lee_2017}.

\subsection{The diffusive approximation}
The assumption of diffusive transport for cosmic rays is relevant if the cosmic rays are travelling through a magnetised medium and when some level of turbulence is present in the magnetic field. This turbulence in the magnetic field prevents the cosmic rays from travelling in straight lines \citep[which would happen, for instance, if no magnetic field was present or if it is sufficiently weak in comparison to the energy of the cosmic rays, such as in][]{rab_2017}. It also means that the cosmic rays are not entirely tied to magnetic fields \citep[such as found in][]{fraschetti_2018}. Instead they scatter off perturbations in the magnetic field structure leading to the cosmic rays diffusing through the system. 

In a spherically symmetric system, and if the cosmic ray energy losses are negligible, then the analytic result for diffusive transport gives a $1/r$ profile. In comparison ballistic propagation, without energy losses, gives a $1/r^2$ profile. Therefore, assuming the same density of cosmic rays at 1\,au for instance, means that if the cosmic rays are diffusing then at 100\,au the density of cosmic rays will be 100 times larger than if they move ballistically. If energy losses become dominant then the radial profile of cosmic rays will deviate from the $1/r$ profile. Fig.\,5.6\ of \citet{longair_2011} shows the rate at which cosmic rays of different energies loss energy as they travel ballistically through different media. In the context of diffusive transport it is important to remember that the cosmic rays are random walking through the disk. As a result the vertical or radial column density of protoplanetary disks does not relate to the path that the cosmic rays have taken through the disk, and therefore the amount of material that they have passed through, to travel out to any given radius. Thus, the column density of the disk is no longer a meaningful quantity to consider. Therefore, we consider the ionisation rates resulting from stellar cosmic rays as a function of radius instead.

\subsubsection{Observations of magnetic fields in protoplanetary disks}
Observationally measuring either the magnetic field structure or its strength in protoplanetary disks has proved immensely challenging. \citet{donati_2005} reported the direct detection of the magnetic field in the core of the protostellar accretion disk of FU Orionis with the high-resolution spectropolarimeter ESPaDOnS. The magnetic field strength reaches strengths of $\sim 1$\,kG close to the centre of the disk. This is obviously a very strong magnetic field, and is partially why FU Orionis was chosen as a target, but FU Orionis is not a typical classical T-Tauri star (CTTS). Nonetheless, this detection supports the idea that magnetic fields are present in protoplanetary disks. If the low energy cosmic rays are ejected outside of the young star's magnetosphere, thus ensuring their escape, it is reasonable to assume that some level of turbulence exists in the magnetic field threading the protoplanetary disk and thus it is valid to assume diffusive transport.

\subsection{Initial conditions}
\label{subsec:init_c}
We expand our previous model so that we can study spatially larger disks, comparable in size to the best studied ALMA samples \citep{ansdell_2016,tazzari_2017}, with a large range of disk masses and surface density profiles. We implemented a logarithmic grid to be able to extend the computational grid in both the $r$ and $z$ direction to $\sim 100$\,au. The number of cells in both the radial and vertical direction is $N_\mathrm{r}=N_\mathrm{z}=60$ giving sufficient spatial resolution for numerically converged results (a convergence test is given in Appendix\,\ref{appendix:a}). At the same time this relatively small number of cells allows us to run many simulations varying the disk mass and the radial power law index which are described in the following sections.

To avoid imposing a small timestep constraint when using the logarithmic grid we set the inner edge of the computational domain to be $R_\mathrm{min}=z_\mathrm{min}=0.1\,$au. Previously the injection site of low energy cosmic rays was taken to be the magnetospheric truncation radius, $R\sim0.07\,$au at a height of $z=0.03\,$au above the disk midplane. Now since $R_\mathrm{min}=z_\mathrm{min}=0.1\,$au we inject the cosmic rays at $(r,z)=(0.14\mathrm{au},0.14\mathrm{au})$. We checked the effect of changing the injection position of the cosmic rays and found it had little effect. It is also important to note that by using $R_\mathrm{min}=z_\mathrm{min}=0.1\,$au we are limited to examining the effect of the cosmic rays at heights $\gtrsim0.1$\,au above the disk midplane.

The outer radial and vertical boundary conditions are absorptive which allow the cosmic rays to diffuse out of the system. The inner radial and vertical boundary conditions are reflective since we assume symmetry about the disk midplane and axial symmetry about the axis of rotation of the disk. The following sections describe the range of values for the physical parameters that we consider for the low energy stellar cosmic rays and for the disk.

\subsection{Physical parameters}
\label{subsec:parameters}
In this section we describe the properties that we use for the low energy cosmic rays. We also outline how we vary the disk mass and the radial density profile, motivated by current observations, in order to investigate the ionising influence of stellar cosmic rays in protoplanetary disks.

\subsubsection{Energy and diffusion coefficient of the cosmic rays}
\label{subsubsec:lcr}
We assume 3\,GeV cosmic rays (protons) and a diffusion coefficient of $D/c = 30r_\mathrm{L}$, where $r_\mathrm{L}$ is the Larmor radius of the particle. We have assumed a mG magnetic field in the calculation of the Larmor radius. This value for the diffusion coefficient is the same as for the fiducial case presented in \citet{rodgers-lee_2017}. The transport properties of low energy cosmic rays in MRI turbulence are unknown and so we can only constrain the diffusion coefficient to be above the Bohm limit, i.e. $D/c > r_\mathrm{L}$ where the diffusion approximation is valid. While we only investigate here the influence of 3\,GeV low energy cosmic rays we still choose a diffusion cofficient for a 3\,GeV proton such that the diffusion coefficient for MeV protons remains above the Bohm limit, since the diffusion coefficient scales with momentum \citep[see Eq.\,5 of][]{rodgers-lee_2017}. For simplicity, we assume mono-energetic cosmic rays of 3\,GeV. 

In Section\,\ref{sec:discussion} we briefly discuss the effect of varying the diffusion coefficient on the results presented in Section\,\ref{sec:results} which assume $D/c = 30r_\mathrm{L}$. In Section\,\ref{sec:discussion} we also discuss the effect of including an energy spectrum of cosmic rays based on our previous findings in \citet{rodgers-lee_2017}. The luminosity of the low energy cosmic rays is taken to be $L_\mathrm{CR}=1\times 10^{28}\mathrm{erg\,s^{-1}}$, as in \citet{rodgers-lee_2017}. 

\subsubsection{Disk mass}
\label{subsec:disk_mass}

For our simulations of cosmic ray propagation the gas mass density of the disk (rather than the dust mass density) is the important quantity since the bulk of the disk mass resides in the gas. This is true if we assume the canonical interstellar medium (ISM) value of 100 for the dust to gas mass ratio. However, while there are a large number of disks with measured dust masses, measuring the gas disk mass has proved more difficult than anticipated \citep[for instance, with ALMA CO observations,][]{miotello_2017}. Therefore, we infer a total disk mass by examining the dust disk masses for which there are larger observational datasets available. We assume that the measured dust mass is a fixed fraction (1/100) of the total mass of the disk.

Observations of protoplanetary disks have shown that the dust disk mass for a given stellar mass displays a large dispersion \citep[Fig.\,7 from][for instance]{pascucci_2016}. To calculate a dust disk mass these observations assume that the dust emission is optically thin, isothermal and adopt a particular value for the dust opacity. Nonetheless, variation of these parameters is not sufficient to reproduce the scatter observed for the disk mass versus stellar mass relationship. This scatter is observed across different star-forming regions, as well as within individual regions.

We use the relationship given in \citet{pascucci_2016} for the dust disk mass as a function of stellar mass which we rearrange to express the total disk mass, $M_\mathrm{disk}$, as
\begin{equation}
M_\mathrm{disk} = 3\times 10^{-4}\bigg[10^{1.1\pm0.8}\left(\frac{M_*}{M_\odot}\right)^{1.9} \bigg]M_\odot
\label{eq:Mdisk}
\end{equation}
This relation is derived from observations of the $\sim2$\,Myr old Chamaeleon star-forming region. For much younger or older star-forming regions the relationship would be different. In Eq.\,\ref{eq:Mdisk} we have adopted the maximum dispersion (0.8) found for the observations which is thought to represent real physical scatter. Thus, we can examine the effect of varying $M_\mathrm{disk}$ for a given stellar mass. Taking $M_*=1M_\odot$ we have $M_\mathrm{disk,min} = 6.0\times 10^{-4}M_\odot$ and $M_\mathrm{disk,max} =2.4 \times 10^{-2}M_\odot$ with a mean value of $3.8\times 10^{-3}M_\odot$. 

\subsubsection{Normalising density, $\rho_0$}
\label{subsubsec:rho0}
The disk mass is calculated from the gas density profile of the disk as
\begin{equation}
M_\mathrm{disk} =  2\int \limits_{z_\mathrm{min}}^{z_\mathrm{max}} \int\limits_{R_\mathrm{min}}^{R_\mathrm{out}}  2\pi R\,\rho(R,z)\,dR\,dz
\label{eq:mdisk}
\end{equation}
The factor of 2 in this equation is due to our assumption of symmetry about the disk midplane. Observations of $R_\mathrm{out}$ for the gaseous disk give $R_\mathrm{out} \sim 70-460\,$au \citep[][for 22 disks in the Lupus star-forming region]{ansdell_2018}. A fiducial value for the disk outer radius $R_\mathrm{out}=100$\,au is used for all of the simulations and we also set $z_\mathrm{max}=R_\mathrm{out}$. The inner boundaries, ${R_\mathrm{min}}$ and ${z_\mathrm{min}}$, are set to $0.1\,$au as described in Section\,\ref{subsec:init_c}. It is important to note that ALMA surveys focusing on the gas disks, as mentioned in \citet{ansdell_2018}, are biased toward the highest mass disks and thus it is possible that lower mass disks with smaller radii are common but undetected by ALMA.

The gas density profile in Eq.\,\ref{eq:mdisk} \citep[as in][]{rodgers-lee_2017} is given by
\begin{equation}
\rho(R,z) = \rho_0e^{-R_\mathrm{in}^2/R^2}\left(\frac{R}{R_0} \right)^{-p} e^{-z^2/2H^2} + \rho_\mathrm{ISM}
\label{eq:rho}
\end{equation}
where $\rho_0$ is the normalising density at $R_0=1$\,au at the midplane of the disk. $R_\mathrm{in}=0.07$\,au is the inner edge of the disk which is taken to be the truncation radius for a typical CTTS \citep[the value for $R_\mathrm{in}$ is described in more detail in][]{rodgers-lee_2017}. It is important to note that $R_\mathrm{in}$ is not the same as $R_\mathrm{min}$. As mentioned in Section\,\ref{subsec:init_c}, $R_\mathrm{min}=0.1\,$au slightly larger than $R_\mathrm{in}$ to avoid a small timestep. $\rho_\mathrm{ISM}=3.89\times10^{-24}\mathrm{g\,cm^{-3}}$ is the density of the ISM and $H=c_\mathrm{s}/\Omega$ is the scale height of the disk where $c_\mathrm{s}$ is the sound speed and $\Omega$ is the Keplerian frequency. The temperature profile used for calculating the sound speed, $c_\mathrm{s}$, is given by,
\begin{equation}
T(R) = T_0\left(\frac{R}{R_0}\right)^{-q}
\label{eq:temp}
\end{equation} 
where $T_0 = T(R_0)=280$K and $q=0.5$ for a solar mass star. 

Eqs.\,\ref{eq:mdisk}-\ref{eq:rho} show that $M_\mathrm{disk}$ can be varied most by varying the following two parameters: the radial power law index, $p$, and the normalising density, $\rho_0$ since we have chosen a fiducial value of $R_\mathrm{out}= 100$\,au. The resulting values of $\rho_0$ chosen to give the minimum, mean and maximum disk masses (for the mean value of $p=1.25$ which is discussed in Section\,\ref{subsubsec:p}) given in Section\,\ref{subsec:disk_mass} are $\rho_0=1.0\times 10^{-13},\,6.8\times 10^{-13}$, and $4.35\times 10^{-12} \,\mathrm{g\,cm^{-3}}$, respectively. In the following section we will discuss the range of values that we consider for the radial power law index, $p$.

\subsubsection{Radial power law index}
\label{subsubsec:p}
As mentioned above the radial power law index, $p$, in Eq.\,\ref{eq:rho} is important in determining the density profile of protoplanetary disks. Observations can probe the dust surface density power law index, $\gamma$, \citep[][for instance]{tazzari_2017} more easily than the gas surface density, where the dust surface density profile can be described by
\begin{equation}
\Sigma(R) = \Sigma_0 \left(\frac{R}{R_0}\right)^{-\gamma}
\label{eq:sigma}
\end{equation}
Integrating over $z$ in Eq.\,\ref{eq:rho} (and assuming a power law distribution for the temperature profile for the disk which only depends on radius such as Eq.\,\ref{eq:temp}) links the power law indices in $\Sigma(R)$, $T(R)$ and $\rho(R,z)$ such that $\gamma = p-3/2+q/2$.

\citet{tazzari_2017} observationally constrain $\gamma$  for 22 (dust) disks around low-mass stars ($\sim 0.1 - 2.0 M_\odot$) in the Lupus star-forming region from their ALMA survey at 890$\mu$m. Instead of using Eq.\,\ref{eq:sigma} exactly they in fact also include an exponential cut-off in the disk surface density characterised by $R_\mathrm{c}$, as in their Eq.\,1. However, if $R_\mathrm{c} > R_\mathrm{out}$ then these equations are very similar. 

\citet{tazzari_2017} find a distribution of values for $\gamma$ centred around $\gamma=0$ with a standard deviation of $0.6$. For our simulations we look at the gas density and therefore, lacking better information, we assume that the gas and dust have the same density profiles. Thus, $-0.6\leq \gamma \leq 0.6$, motivated by observations, gives $0.65\leq p \leq 1.85$, assuming $q = 0.5$. For the MMSN model $\gamma=1.5$ and $p=2.75$, again assuming $q=0.5$. Thus, we investigate $0.65\leq p \leq 2.75$ to extend the range to include the values used in the MMSN model. In \citet{rodgers-lee_2017} we investigated $0.5\leq p \leq 1.5$ which are similar to the values inferred from observations.

\subsection{The simulations}
We study the effect of varying the disk mass and of varying $p$ via a parameter space survey.  We choose 12 logarithmically spaced disk masses in the range $M_\mathrm{disk} = 6 \times 10^{-4}-2.4 \times 10^{-2} M_{\odot}$.  For each of these 12 disk masses we perform simulations for 15 linearly spaced values of $p$ in the range 0.65 -- 2.75 (described above).  The results presented in the next section are based, therefore, on a suite of 180 separate simulations.

\section{Results}
\label{sec:results}
In this section we present the results from our simulations. In Sections\,\ref{subsec:fid}-\ref{subsec:mmsn} we will first begin by discussing a number of specific simulations in detail. Namely, the results relating to the mean, minimum and maximum disk mass simulations. We will also present results for the mean disk mass for varying density profiles. We will discuss the results obtained for the disk parameters most similar to those of the MMSN. Finally, we present a more general view of the results in Section\,\ref{subsec:100au}. 

The main diagnostic that we examine is the ionisation rate, $\zeta_\mathrm{CR}$, resulting from low energy stellar cosmic rays. The ionisation rate is calculated using Eq.\,15 from \citet{rodgers-lee_2017} which is based on Eq.\,21 of \citet{umebayashi_1981}. It is important to note that, as discussed in Section\,\ref{subsec:init_c}, the smallest vertical height above the disk midplane that we can examine the ionisation rate at is $\sim0.14$\,au. In comparison, in \citet{rodgers-lee_2017} we were able to investigate a smaller vertical height above the disk midplane of $0.03$\,au. At small radii ($\lesssim0.5$\,au) the vertical density profile changes rapidly since the scale height of the disk is small. Thus, the ionisation rate at 0.14\,au above the disk midplane may differ from the ionisation rate at the disk midplane. In Fig.\,4 of \citet{rodgers-lee_2017} the black and blue dots show the ionisation rate as a function of radius at a height of 0.03 and 0.1\,au above the disk midplane, respectively. For radii $<0.1\,$au the ionisation rate differs by approximately a factor of two for the different vertical heights above the disk midplane. At 1\,au, there is approximately a factor of three difference between the ionisation rates but this gradually decreases for larger radii. At large radii the change in the density of the disk at the midplane and at 0.14\,au above the disk midplane will be relatively small. Therefore, at large radii examining the ionisation rate at 0.14\,au above the disk midplane will be a good estimate of the ionisation rate at the disk midplane.

\subsection{Mean $M_\mathrm{disk}$ and mean value of $p$}
\label{subsec:fid}
\begin{figure}
        \includegraphics[width=0.5\textwidth]{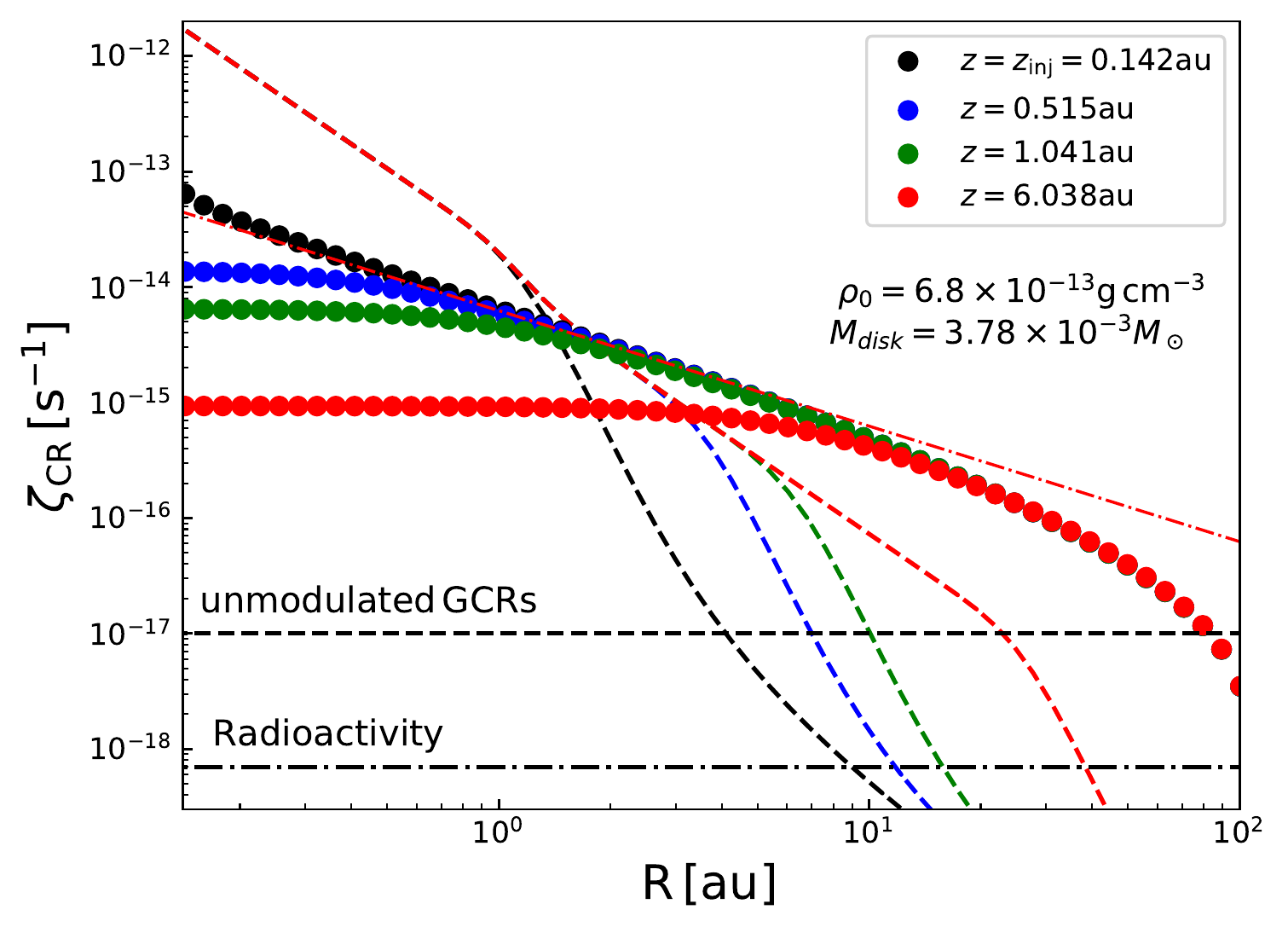}
\caption{This plot shows the ionisation rate due to 3\,GeV low energy cosmic rays as a function of radius for varying heights above the midplane of the disk. The disk mass for this simulation is the mean disk mass described in Section\,\ref{subsec:disk_mass} and the radial density profile has $p=1.25$ in Eq.\,\ref{eq:rho}.  The assumed value of $\rho_0$ is indicated on the plot. The red dash-dotted line indicates the analytic result for spherically symmetric diffusion, without considering any energy losses. The ionisation rate expected from unmodulated galactic cosmic rays and also from radioactivity are shown on the plot for comparison. The dashed lines represent the ionisation rate from stellar X-rays for the same heights above the disk midplane as for the stellar cosmic rays.\label{fig:meanMdisk}}
\end{figure}

We start by examining the ionisation rate from low energy cosmic rays obtained for the simulation with the most typical disk parameters (motivated by observations): the mean disk mass ($M_\mathrm{disk}=3.78\times10^{-3}M_\odot$) and the mean value of $p=1.25$. These values result in $\rho_0=6.8\times10^{-13}\mathrm{g\,cm^{-3}}$. Fig.\,\ref{fig:meanMdisk} shows the resulting ionisation rate as a function of radius in the disk for different vertical heights above the disk midplane. For reference, the ionisation rate expected from unmodulated galactic cosmic rays \citep[$\zeta_\mathrm{GCR}\sim10^{-17}\,\mathrm{s^{-1}}$,][]{umebayashi_1981} and radioactivity \citep[$7\times10^{-19}\,\mathrm{s}^{-1}$,][]{umebayashi_2009} are given by the black dashed and dash-dotted lines, respectively. The dashed lines represent the ionisation rate from stellar X-rays for different vertical height above the disk midplane, using the same parameterised fit from \citet{bai_2009}, as used in \citet{rodgers-lee_2017}. The assumed X-ray luminosity is $10^{29}\mathrm{erg\,s^{-1}}.$

The first most noticeable result in Fig.\,\ref{fig:meanMdisk} is that the $1/r$ profile expected from spherically symmetric spatial diffusion (in the absence of losses) is recovered out to $\sim10$\,au at the disk midplane, as indicated by the black points. The red dash-dotted line indicates the analytic result for spherically symmetric diffusion, without considering any energy losses. The black points trace the ionisation rate at a vertical height of $0.142$\,au above the disk midplane. This result is significantly different from our previous findings in \citet{rodgers-lee_2017} where we found that the cosmic rays were strongly attenuated within $\sim1$\,au and had a much steeper than $1/r$ profile. We discuss in detail the reasons for this difference in Section\,\ref{sec:discussion}. Essentially though this difference merely represents a difference in the underlying assumed density profile of the disk.

\begin{figure}
        \includegraphics[width=0.5\textwidth]{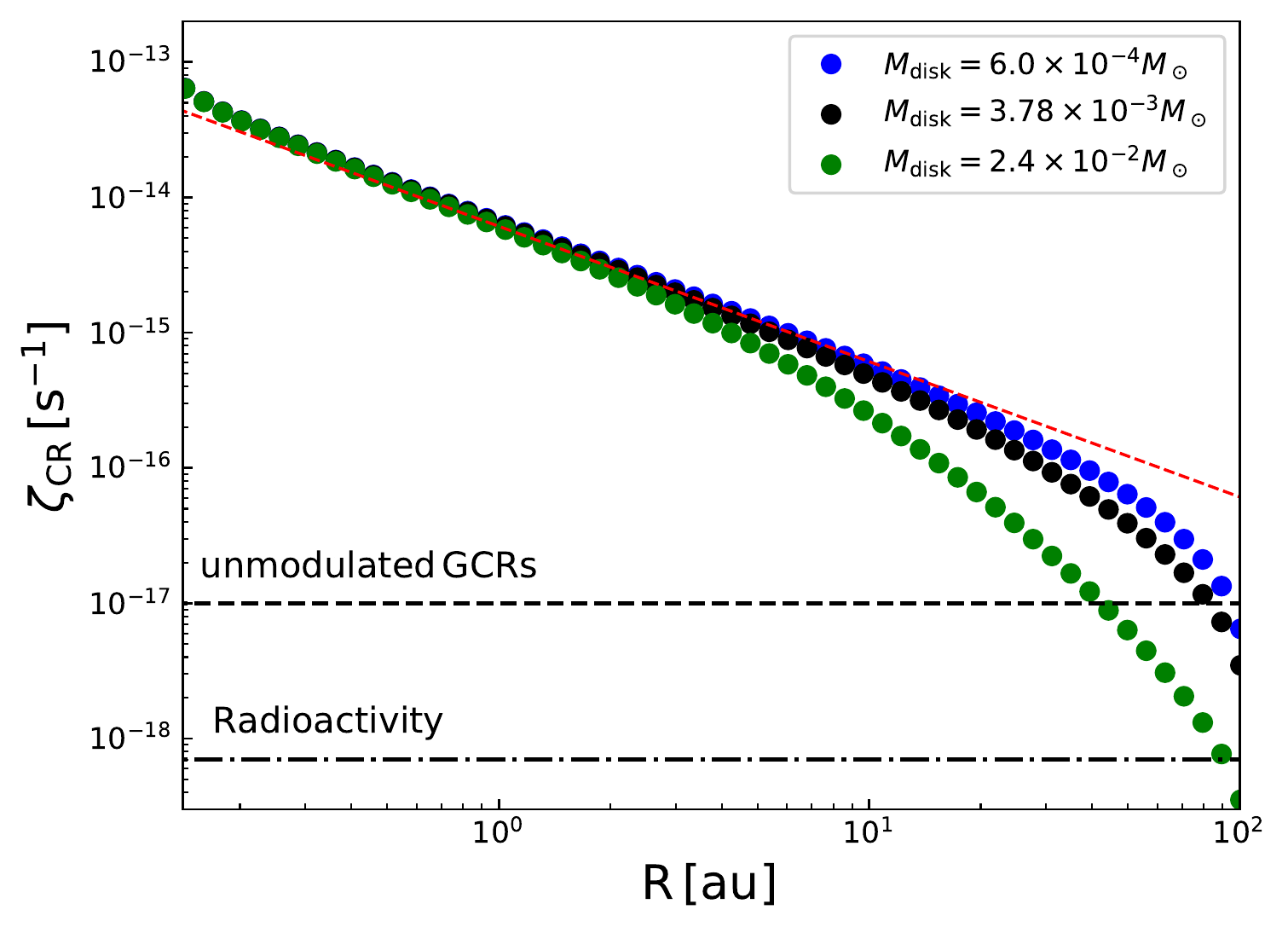}
\caption{This plot shows the ionisation rate as a function of radius at $z=0.142\,$au above the midplane of the disk for the minimum, mean and maximum disk masses examined here using the mean value of $p=1.25$ for all three cases.}
    \label{fig:min_maxMd}
\end{figure}

The second interesting result is that the ionisation rate resulting from low energy stellar cosmic rays is larger than the ionisation rate expected from unmodulated galactic cosmic rays out to $\sim80$\,au at a vertical height of $0.142$\,au above the disk midplane. The ionisation rate from stellar cosmic rays also dominates over the contribution from stellar X-rays beyond $\sim4$\,au. This indicates that for this `typical' disk low stellar energy cosmic rays can result in high levels of ionisation deep within the disk out to $\sim80$\,au.

The blue, green and red points show the ionisation rate at increasing heights above the disk midplane in Fig.\,\ref{fig:meanMdisk}. As expected, at small radii but high above the disk midplane the ionisation rate is smaller than for lower heights. For $R\gtrsim10\,$au the ionisation rate is the same irrespective of the height above the disk midplane. The lower ionisation rates displayed at large vertical heights (the red dots out to $\sim 10$\,au for instance) occur at large values of $z/R$. This merely reflects that the spherical radius, is then significantly larger the cylindrical radius at these heights and radii. The ionisation rate is in fact nearly spherically symmetric which implies that this particular disk is, for the most part, `optically thin' to the cosmic rays.

Another interesting quantity to examine is the ionisation fraction which can be used to determine the level of coupling between the neutral and charged species. This level of coupling is important to determine for non-ideal magneto-hydrodynamic simulations of protoplanetary disks. Based on the simplified chemical network from \citet{fromang_2002} we can estimate the ionisation fraction using Eqs.\,16-17 of \citet{rodgers-lee_2017} by choosing an appropriate recombination rate coefficient. 

We consider two extreme cases which represent when metals are present or absent in the gas. If the metal ions are locked in sedimented grains we calculate the ionisation fraction at $z=0.142\,$au above the disk midplane to be $6.3\times10^{-9},\,1.2\times10^{-10},\,1.1\times 10^{-10}$ and $7.8\times 10^{-11}$ at $R=1, 10, 30$ and $70$\,au, respectively. If metal ions instead dominate as charge carriers in the plasma then we find that the ionisation fraction at $z=0.142\,$au above the disk midplane to be $1.8\times10^{-6},\,3.9\times10^{-8},\,3.5\times 10^{-8}$ and $2.5\times 10^{-8}$ at $R=1, 10, 30$ and $70$\,au, respectively. 

The ratio of neutral-ion collision time in comparison to the orbital time gives an estimate of the level of coupling between the ions and the neutral gas which can be used as a criteria to determine whether the MRI will be active or not in the disk \citep[Eq.\,2 from][for instance]{chiang_2007}. This ratio depends on the ionisation fraction in the disk. For the ionisation fractions given above we find that if metal ions are locked in sedimented grains the neutral-ion coupling is poor. On the other hand, we find, if metal ions dominate as charge carriers, that the ionisation fractions recovered due to stellar cosmic rays may be important for the dynamics of protoplanetary disks since the neutral-ion collision time is sufficiently small in comparison to the orbital time.

\subsection{Minimum and maximum values of $M_\mathrm{disk}$ with $p=1.25$}
\label{subsec:vary_md}
Next, we examine the effect of varying the disk mass from the minimum ($M_\mathrm{disk}=6\times10^{-4}M_\odot$) to the maximum ($M_\mathrm{disk}=2.4\times10^{-2}M_\odot$) values that we motivated in Section\,\ref{sec:form}. We keep $p=1.25$ fixed here. The resulting ionisation rate profiles for the minimum, mean and maximum disk masses are shown in Fig.\,\ref{fig:min_maxMd}. For the minimum disk mass simulation the ionisation due to low energy cosmic rays now results in $\zeta_\mathrm{CR}>10^{-17}\mathrm{s^{-1}}$ effectively throughout the whole disk (blue dots in Fig.\,\ref{fig:min_maxMd}). On the other hand, as expected, for the maximum disk mass case the stellar cosmic rays are more attenuated, such that at $\sim2$\,au the ionisation rate profile is steeper than $1/r$ (green dots in Fig.\,\ref{fig:min_maxMd}). Nonetheless, for the typical most massive disk that we consider the low energy cosmic rays still out-compete the unmodulated galactic cosmic ray ionisation rate out to a radius of $\sim40$\,au for all heights above the disk. Beyond $R\gtrsim20$\,au at the disk midplane, the difference in the resulting ionisation rate between the most massive disk and the least massive disk is approximately an order of magnitude.

\subsection{Mean disk mass with varying density profiles }
\label{subsec:vary_p}
In this section we focus on the effect of varying the density profile of the disk while keeping the disk mass fixed as the mean disk mass. We varied $p=0.65-2.75$, where the range of $p=0.65-1.85$ is motivated by observations and the additional values up to $p=2.75$ reflects our interest in considering disks with parameters similar to those of the MMSN disk which has $p=2.75$.

\begin{figure}
        \includegraphics[width=0.5\textwidth]{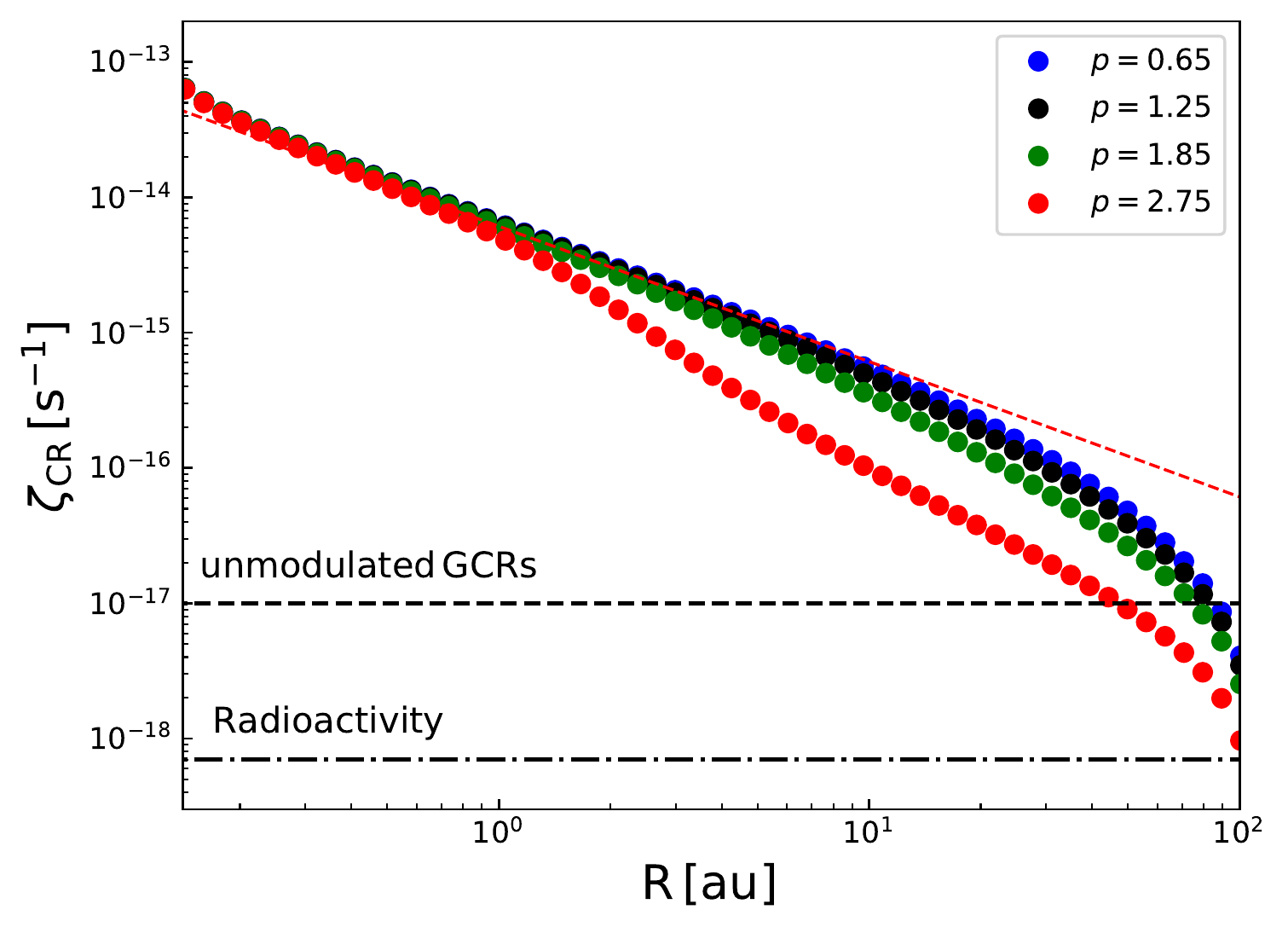}
  \caption{This plot shows the effect of varying the density profile on the resulting ionisation rate for the mean disk mass ($M_\mathrm{disk}=3.78\times 10^{-3}\,M_\odot$). The ionisation rate is given a function of radius at $z=0.142\,$au above the midplane of the disk.} 
    \label{fig:vary_p}
\end{figure}

Fig.\,\ref{fig:vary_p} shows the radial ionisation rate profiles considering the mean disk mass with $p=0.65-2.75$. For $p=0.65$ and $p=1.85$ (the blue and green dots in Fig.\,\ref{fig:vary_p}) the corresponding radial ionisation rate profiles are very similar to the profile obtained using the mean value of $p=1.25$ (the black dots). Generally though for a fixed disk mass, steeper disk density profiles result in the stellar cosmic rays being attenuated more. For the most extreme case, corresponding to the MMSN value of $p=2.75$, $\zeta_\mathrm{CR}>10^{-17}\mathrm{s^{-1}}$ only out to $R\sim 50$\,au. The radial profile, as shown by the red dots in Fig.\,\ref{fig:vary_p}, deviates from the $1/r$ profile within $R\sim$1\,au. Therefore, as will be discussed in more detail in Section\,\ref{subsec:100au}, massive disks combined with steep density profiles will be very effective in impeding the transport of low energy cosmic rays far out in the disk. The important point to note is that for the mean disk mass and the observed range of values of $p$ considered, therefore not considering the (extreme) MMSN value of $p$, the cosmic rays are not significantly attenuated by the disk.

\subsection{MMSN-type disk}
\label{subsec:mmsn}
The MMSN is a model which defines the minimum mass and distribution of material as a function of radius in a disk required to reproduce the distribution of planets in the solar system in their current locations \citep{hayashi_1981}. This model assumes that the planets in the solar system formed at their current locations without migrating. The resulting parameters are that the radial extent of the protosolar disk was $R=0.35-36$\,au with $\rho_0 = 1.4 \times 10^{-9}\,\mathrm{g\,cm^{-3}}$ and $p=2.75$ which gives $M_\mathrm{disk}=0.013M_\odot$. 

\begin{figure}
	\centering
        \includegraphics[width=0.5\textwidth]{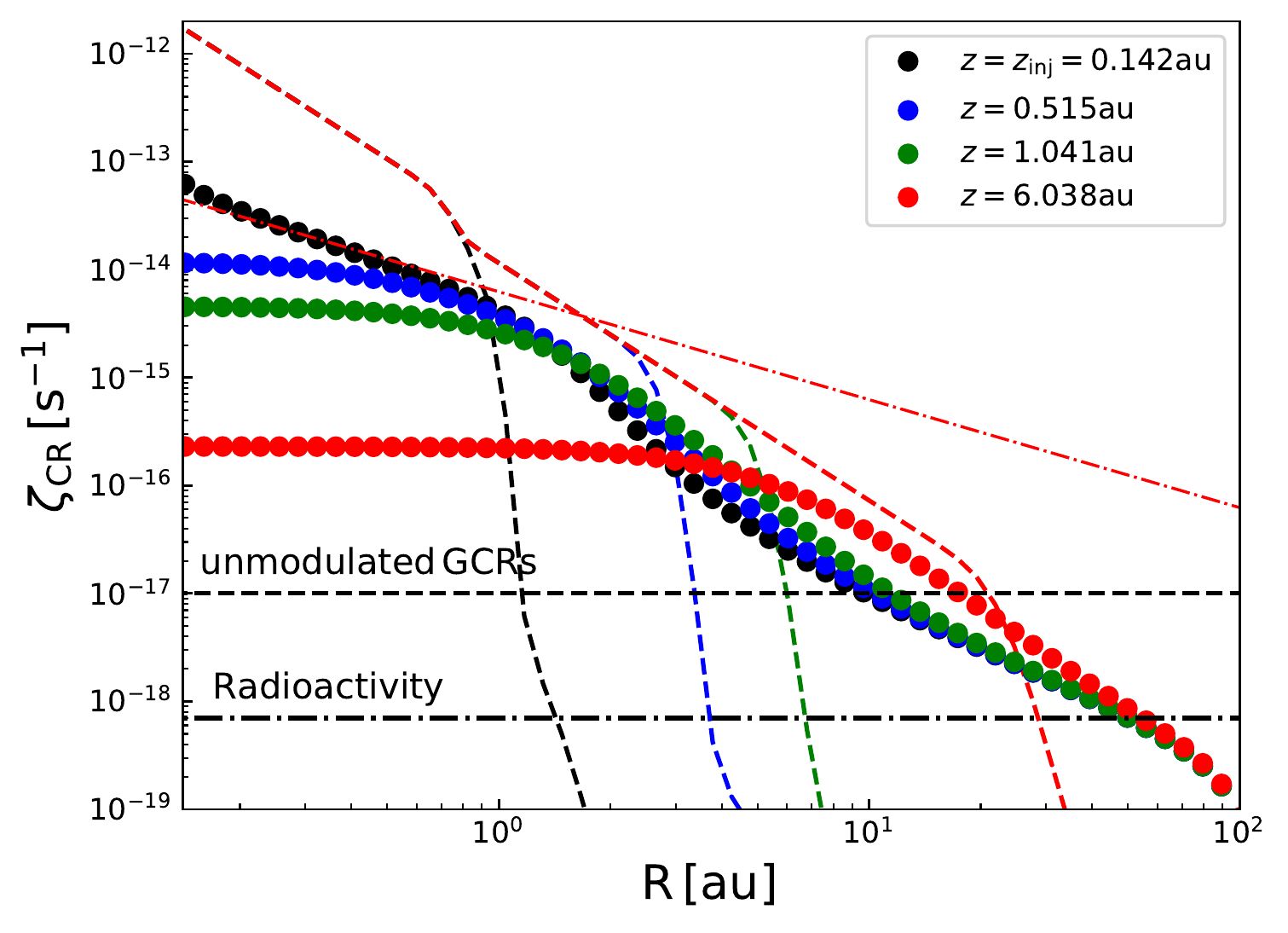}
       	\centering
  \caption{Radial ionisation rate profile for a MMSN-type disk which has $p=2.75$ and $\rho_0\sim1.4\times10^{-9}\,\mathrm{g\,cm^{-3}}.$} 
    \label{fig:mmsn}
\end{figure}

The simulation that we refer to as a `MMSN-type' disk has $\rho_0 = 1.4 \times 10^{-9}\,\mathrm{g\,cm^{-3}}$ and $p=2.75$ but retains the radial extent that we use for all of our simulations, $R=0.1-100$\,au. This means $M_\mathrm{disk}=0.024M_\odot$ which is at the upper end of the observed disk mass distribution for solar mass stars. This steep density profile combined with the large disk mass means that the low energy cosmic rays are unable to travel further than $\sim10$\,au in the disk, shown in Fig.\,\ref{fig:mmsn}. A protoplanetary disk with these parameters would be an observational outlier. This MMSN-type disk has the same mass as the simulation shown by the green dots in Fig.\,\ref{fig:min_maxMd}. By comparing with the ionisation rate given by the black dots in Fig.\,\ref{fig:mmsn} it is clear that the value of $p$ for disks this massive has a large effect. The dashed lines in Fig.\,\ref{fig:mmsn} represent the X-ray ionisation rate, as in Fig.\,\ref{fig:meanMdisk}, for the same vertical heights above the disk midplane as shown for the stellar cosmic rays. For this MMSN-type disk, the stellar X-rays dominate as the source of ionisation out to 30\,au. The low energy cosmic rays dominate in the very outer regions of this disk but still result in a very low ionisation rate.

\subsection{Global trends: The ionisation rate due to stellar cosmic rays}
\label{subsec:100au}
The previous sections have focused on a number of individual simulations to give some insight into the effect of varying the disk mass and the disk radial profile. We now examine the results obtained by considering the overall suite of simulations.  This allows us to infer the influence of stellar cosmic rays on the global population of observed protoplanetary disks around solar-mass stars.  We perform this analysis by examining the ionisation rate due to stellar cosmic rays at $z = 0.142$\,au ($\sim$ disk midplane) and identify whether $\zeta > \zeta_{\rm GCR} = 10^{-17}$\,s$^{-1}$ for each simulation. We make this comparison at a number of different radii.

Over the whole range of disk parameters we have explored, stellar cosmic rays dominate the midplane ionisation ($\zeta_\mathrm{CR}>10^{-17}\,\mathrm{s^{-1}}$) within the first 10\,au of the disk. This means that irrespective of the disk density profile or the disk mass (for the observationally motivated ranges we investigated) that low energy cosmic rays provide an important source of ionisation out to 10\,au at the midplane of the disk.

\begin{figure}
        \includegraphics[width=0.5\textwidth]{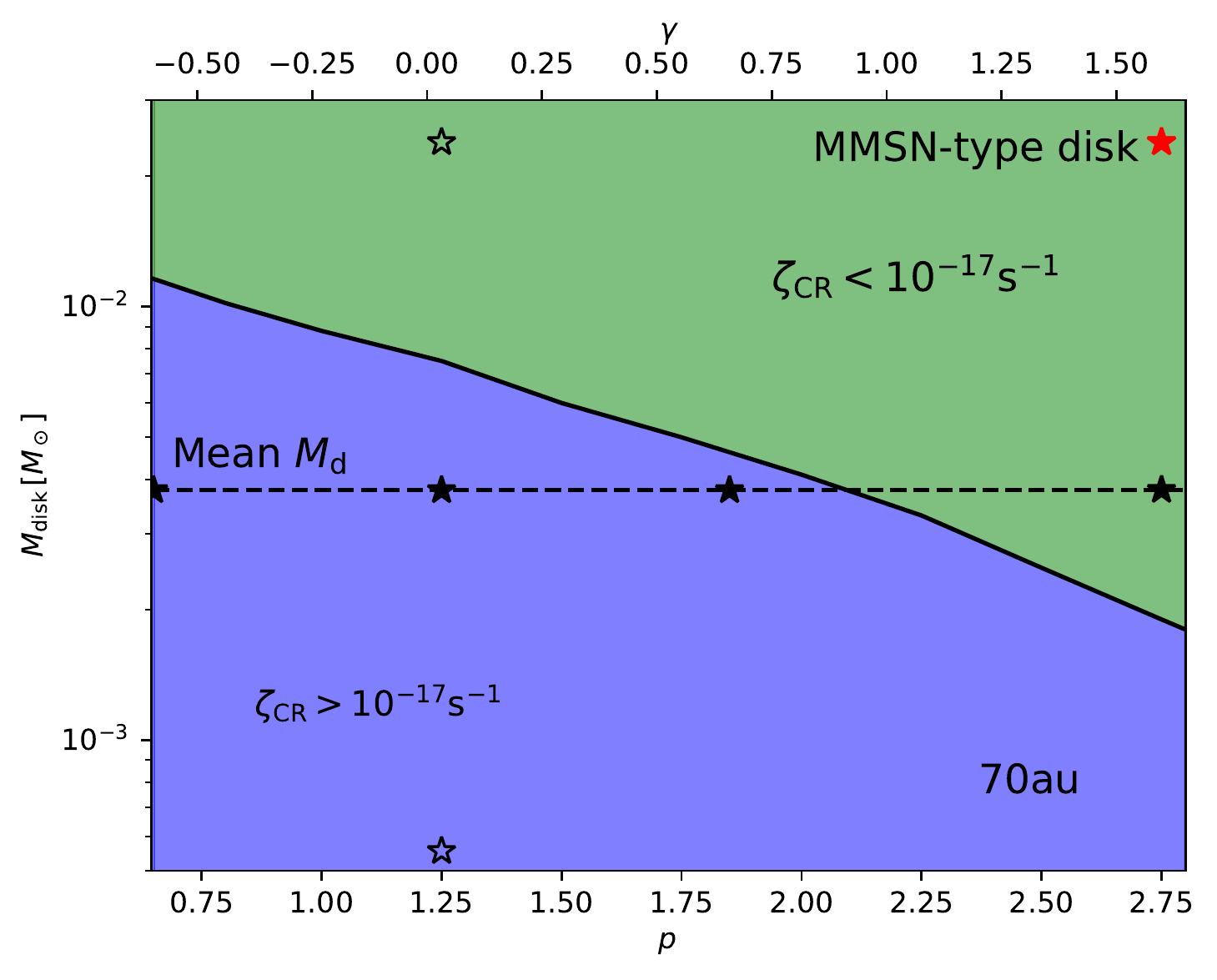}
\caption{This plot shows two shaded regions as a function of $M_\mathrm{disk}$ and $p$. The blue shaded region represents simulations with values of $M_\mathrm{disk}$ and $p$ for which the ionisation rate from low energy stellar cosmic rays dominates over that expected from unmodulated galactic cosmic rays ($\zeta_\mathrm{CR}>10^{-17}\mathrm{s^{-1}}$) at $R=70$\,au at the midplane of the disk. The green shaded region corresponds to those simulations where $\zeta_\mathrm{CR}<10^{-17}\mathrm{s^{-1}}$. The dashed line shows the mean disk mass value ($3.78\times 10^{-3}M_\odot$). The black stars represent the simulations discussed in Sections\,\ref{subsec:fid}-\ref{subsec:vary_p} and the red star represents the MMSN-type disk investigated in Section\,\ref{subsec:mmsn}. 
\label{fig:Md_p}}
\end{figure}

However, for a large number of disks the region where stellar cosmic rays dominate extends much further. For example, we determine that $\sim$50\% of the disks have $\zeta_\mathrm{CR}>10^{-17}\,\mathrm{s^{-1}}$ at the midplane of the disk at $\sim$70\,au. Fig.\,\ref{fig:Md_p} shows the results of this comparison for $R=70$\,au. The blue shaded region indicates the region of parameter space for which the ionisation rate from stellar cosmic rays is greater than $10^{-17}\,\mathrm{s^{-1}}$ and the green shaded region shows where it is less than it. 

The horizontal black dashed line in Fig.\,\ref{fig:Md_p} denotes the mean disk mass for the Chamaeleon star-forming region, as discussed in Section\,\ref{sec:form}. The filled black stars represent the simulations discussed in Section\,\ref{subsec:vary_p} which vary $p$ for the mean disk mass. The unfilled stars represent the simulations discussed in Section\,\ref{subsec:vary_md} which vary the disk mass for the mean value of $p=1.25$. The red star denotes the MMSN-type disk simulation with $\rho_0\sim1.4\times10^{-9}\mathrm{g\,cm^{-3}}$ and $p=2.75$ (as described in Section\,\ref{subsec:mmsn}). 

Fig.\,\ref{fig:Md_p} shows that for disk masses between $M_\mathrm{disk}\sim2\times10^{-3}-1\times10^{-2}M_\odot$ at a radius of 70\,au the density profile of the disk is an important quantity. Variations in $p$ result in the ionisation rate due to stellar cosmic rays becoming bigger or smaller in comparison to the ionisation rate due to unmodulated galactic cosmic rays at the disk midplane. Interestingly, the mean disk mass observed in Chamaeleon for a solar mass star, shown by the black dashed line in Fig.\,\ref{fig:Md_p} lies within this region of parameter space. Thus, the level of ionisation present in many disks will be very dependent on the density profile of the disk.

For more massive disks with $M_\mathrm{disk}>1\times10^{-2}M_\odot$ the ionisation rate due to stellar cosmic rays is $<1\times 10^{-17}\mathrm{s^{-1}}$ irrespective of the density profile of the disk at 70\,au. Conversely for less massive disks with $M_\mathrm{disk}<2\times10^{-3}M_\odot$ stellar cosmic rays dominate irrespective of the density profile of the disk for the values of $p$ that we consider. For radii larger than 70\,au the number of simulations with $\zeta_\mathrm{CR}>10^{-17}\,\mathrm{s}^{-1}$ decreases but remains significant, $\sim$25\%  at 90\,au for instance. The slope of the dividing line remains quite similar and shifts downwards. 

\section{Discussion}
\label{sec:discussion}
We have performed a parameter space study of the disk mass and radial profile of the gas density to investigate the effect they have on the ionisation from low energy stellar cosmic rays. We have focused on the simplest case with mono-energetic low energy cosmic rays assuming a value for the diffusion coefficient. Therefore in this section we discuss the possible effect of including an energy spectrum of cosmic rays and the effect of varying the diffusion coefficient. We also compare with our previous results \citep{rodgers-lee_2017} and to other results in the literature.

\subsection{Effect of varying the diffusion coefficient}
In Section\,\ref{sec:results} we have assumed a diffusion coefficient of $D/c = 30r_\mathrm{L}$ for 3\,GeV protons. Fig.\,6 in \citet{rodgers-lee_2017} gives an overall picture of the effect of varying the diffusion coefficient for a very dense disk. As can be seen from this plot, increasing the diffusion coefficient results in a decrease in the ionisation rate at small radii. At the same time the radial profile of the ionisation rate approaches a $1/r$ profile at larger radii for larger diffusion coefficients as the cosmic rays spend less time in the dense, and therefore more attenuating, inner region of the disk. In the absence of significant losses changing the diffusion coefficient would simply result in a vertical shift in the ionisation profile with a $1/r$ profile throughout the system. 

Relating to the results presented here the first point to note is that the effect of changing the diffusion coefficient will depend on the density profile of the disk. We can start with any of the simulations that resulted in a $1/r$ profile (such as Fig.\,\ref{fig:meanMdisk} and many of the simulations in the blue shaded region of Fig.\,\ref{fig:Md_p}). We can estimate for these disks that the resulting change in the radial ionisation rate obtained by increasing the diffusion coefficient will be a scaling law, since the steady state ionisation rate for cases in which the diffusive escape time is shorter than the ionisation loss time is $\zeta_\mathrm{CR} \sim 1/Dr$. Larger diffusion coefficients than the value adopted for these cases are therefore expected to lead to a proportionally smaller steady state ionisation rate.

The effect of decreasing the diffusion coefficient may not be as simple. If the diffusion time becomes significantly longer than the ionisation loss timescale then behaviour similar to that seen in Fig.\,6 in \citet{rodgers-lee_2017} may be observed. There the cosmic rays are absorbed faster than they can escape the dense inner regions of that disk which results in profiles steeper than $1/r$.

We can also consider the simulations occupying the green shaded parameter space in Fig.\,\ref{fig:Md_p} (such as the MMSN-type disk in Fig. \ref{fig:mmsn}) which already indicate, to some extent, radial ionisation rate profiles steeper than $1/r$. It is possible for some of the disks in this parameter regime that increasing the diffusion coefficient will decrease the diffusion time sufficiently such that the low energy cosmic rays will be able to reach the outer regions of the disk (such as the most massive disk simulation, shown as the green dots in Fig.\,\ref{fig:min_maxMd}, since it is not as far from a $1/r$ profile as Fig.\,\ref{fig:mmsn}, for instance). On the other hand, decreasing the diffusion coefficient for disks in this parameter regime will again result in even steeper radial profiles, as mentioned above.

\subsection{Effect of including an energy spectrum of cosmic rays}
We have only investigated mono-energetic stellar cosmic rays in this paper for simplicity. Again, the effect of including an energy spectrum of cosmic rays was investigated in \citet[Fig.\,7,][]{rodgers-lee_2017}. As evidenced by this plot the behaviour that results by including an energy spectrum of cosmic rays is complicated and will depend on the density profile of the disk. We will consider the energy spectrum used in \citet{rodgers-lee_2017} which assumed a spectral index of -2 with stellar cosmic rays of energies from $~\sim 100$\,MeV to 300\,GeV. We can briefly comment that disks which exist in the green shaded region of Fig.\,\ref{fig:Md_p} may become more ionised further out in the disk due to the high energy component present in the energy spectrum.

\subsection{Comparison to previous work}
The main finding in \citet{rodgers-lee_2017} was that low energy cosmic rays were strongly attenuated by the dense protoplanetary disk. They were only competitive with unmodulated galactic cosmic rays as a source of ionisation out to a maximum radius of $\sim 1$\,au. This is in contrast to the behaviour reported in this paper. The important difference is the assumed density profile and normalising density. The disk investigated in \citet{rodgers-lee_2017} had $\rho_0 = 2.33\times 10^{-9}\mathrm{g\,cm^{-3}}$ and $p=1.0$ which results in a very dense disk ($M_\mathrm{disk}\sim 0.1M_\odot$ with a radial extent of 10\,au as considered previously), outside of the parameter ranges that we investigated here. Such a dense and compact disk may exist but would be more representative of a more embedded, and therefore younger, source. On the other hand, some of the disks in Lupus observed in CN with ALMA shown in \citet{van_terwisga_2019} are thought to have very small disk radii, $R\sim15$\,au but none of them are quite as massive as the disk we previously considered. Again, it is worth noting that massive optically thick compact disks may exist and remain undetected in ALMA surveys.

Thus, the parameters we vary here to investigate the ionising influence of low energy cosmic rays are likely to be more representative of the majority of protoplanetary disks. Our results indicate that for 50\% of the simulations which varied the disk masses and density profiles that low energy stellar cosmic rays would be an important source of ionisation out to 70\,au in radius very close to the midplane. We also showed that a MMSN-type disk (with a larger disk mass than the MMSN disk model as a result of the outer radius being extended to 100\,au while retaining the same values for $p$ and $\rho_0$) would be an observational outlier since the disk is quite massive in combination with a very steep density profile. The MMSN-type disk we investigated was very effective at preventing stellar cosmic rays from penetrating deep into the disk at large radii.

\subsection{Comparison to the literature}
\citet{rab_2017} present their stellar cosmic ray energy spectrum as a differential intensity in their Figs.\,1 and 2. This has the advantage of being able to make a direct comparison with the local interstellar spectrum of galactic cosmic rays measured by Voyager \citep{stone_2013} at 122\,au and with the modulated galactic cosmic ray spectrum measured at 1\,au on Earth \citep[\emph{PAMELA} measurements from 2006-2009,][for instance]{adriani_2013}. Their assumed energy spectrum is based on observed solar energy particle spectra. They take one such spectrum from \citet{mewaldt_2005}, divide by the flare duration and multiply it by $10^5$ to obtain their stellar energy particle spectrum. This results in a total energy injection rate of $10^{30} {\rm erg\,s^{-1}}$ with the majority of cosmic rays having MeV energies. In comparison, we inject a power ($L_\mathrm{CR}\sim 10^{28}\mathrm{erg\,s^{-1}}$) of only 3\,GeV protons. There are a number of reasons that we find the low energy cosmic rays to be more effective as a source of ionisation in comparison to \citet{rab_2017}. 

First, the ionisation loss rate for MeV protons is much larger than for GeV protons \citep[see Fig.\,5.6 of][]{longair_2011}. This means, despite the power used in \citet{rab_2017} being two orders of magnitude larger than used here, that the MeV protons are not able to travel as far as GeV protons in the disk as they suffer larger energy losses. Second, their assumption of rectilinear propagation results in a steeper radial profile than the $1/r$ profile we recover for many of our simulations. Finally, the disk density profile and other physical parameters ($M_*=0.7M_\odot$ and $M_\mathrm{disk}=0.01\,M_\odot$, for instance) are different and so it is difficult to make direct comparisons. Nonetheless, the above reasons highlight that the overall power injection of cosmic rays is not necessarily an indication of what the ionisation rate will be and that the energy of the cosmic rays is more important, combined with the assumption of their transport mechanism.

As mentioned in Section\,\ref{sec:intro}, \citet{cleeves-2015} constrain the total ionisation rate for the disk of TW Hya ($\sim 0.8M_\odot$ CTTS) to be $\zeta\lesssim 10^{-19}\mathrm{s^{-1}}$. They find a total gas mass for the disk of $0.04 \pm 0.02M_\odot$ with an assumed value of $\gamma=1$. The density profile also assumes an exponential cut-off radius of 150\,au which is $>100$\,au ($R_\mathrm{out}$ for our disks) and therefore our density profiles should be similar within this radius. It is also important to note that our value of $\gamma=1$ assumes a particular value of $q$, and therefore $p$, which is also different from that used in \citet{cleeves-2015}. Changing $q$ will result in the scale height of the disk changing.

By comparing with the observations presented in \citet{pascucci_2016} it is apparent that TW Hya's disk is relatively massive. The most massive disk that we consider ($2\times10^{-2}M_\odot$) with $\gamma=1$ has $\zeta_\mathrm{CR}\lesssim 10^{-19}\mathrm{s^{-1}}$ for $R>80\,$au for all heights above the disk. If the disk mass were doubled then $\zeta_\mathrm{CR}\lesssim 10^{-19}\mathrm{s^{-1}}$ should occur at smaller radii. It is important to note that we consider a solar mass star, whereas TW Hya has a mass of $\sim 0.8M_\odot$. This will alter the above estimate by changing the scale height of the disk. Nonetheless, the observational constraint on the ionisation rate for TW Hya is not inconsistent with our results and as we have shown the ionisation rate depends sensitively on the disk mass. 

\citet{fraschetti_2018} perform test particle numerical simulations of the transport of $\sim$GeV stellar cosmic rays through the wind of a T Tauri star. They find that the ionising effect of the cosmic rays occurs in only localised regions of the disk (specifically when they are the dominant source of ionisation in comparison to X-rays) and close to the region of injection. \citet{fraschetti_2018} inject the cosmic rays at various radii between $2-10\,R_*(=0.009-0.04\,\mathrm{au})$ which is within the truncation radius of the disk. In their case the cosmic rays propagate until their trajectories intersect with the disk surface (and lose all their energy at the interaction point) or take them back to the stellar surface. This would be equivalent to assuming a very large density for the disk in our case which is why our earlier results in \citet{rodgers-lee_2017} appeared consistent with their findings.

The work of \citet{fraschetti_2018} highlights that the ionisation rate is possibly not symmetric around the axis of rotation of the star. By comparing with our results it suggests that the cosmic rays need to be injected outside of the truncation radius in order to be able to significantly ionise the disk further out.

\section{Conclusions}
\label{sec:conclusions}
In this paper we have assessed the ionising effect of low energy cosmic rays in protoplanetary disks originating close to the central star while varying the disk mass and the radial density profile of the disk. The variation in these parameters was motivated by current observations of protoplanetary disks \citep{pascucci_2016, tazzari_2017}.

We found that the $1/r$ profile expected for spherically symmetric diffusion, in the absence of significant energy losses, was recovered for many of the simulations in the parameter space that we investigated. This is in contrast to the results presented in \citet{rodgers-lee_2017} because of the assumed normalising density at 1\,au. Effectively, we previously investigated a dense and compact disk which is likely to be an outlier for protoplanetary disk populations. Whereas, in this paper we have focused on values around the mean observed disk mass.

We found that the low energy stellar cosmic rays provide an ionisation rate greater than expected from unmodulated and unattenuated galactic cosmic rays out to a radius of $\sim$10\,au near the midplane of the disk for all the simulations. For at least 50\% of the simulations the low energy cosmic rays continued to effectively ionise the midplane of the disk out to a radius of $\sim 70$\,au. 

The mean disk mass, taken from dust observations of protoplanetary disks in the Chamaeleon star-forming region, combined with the mean value for the radial dust density profile index, lies within the parameter space of disks that display high levels of ionisation, relative to that expected from unmodulated galactic cosmic rays, at 70\,au. In comparison, the maximum disk mass (within the $1\sigma$ scatter expected from observations) results in the ionisation rate decreasing by approximately an order of magnitude at 70\,au.

The MMSN-type disk that we investigated (with a radial extent of 100\,au instead of 36\,au which results in $M_\mathrm{disk}\sim0.024M_\odot$) has a steep disk density profile and is quite massive in comparison to observed protoplanetary disks. This combination of parameters meant that this MMSN-type disk is one of the most effective disks at excluding low energy cosmic rays from the outer regions of the disk. 

Our results are consistent, within the uncertainties of our model parameters, with the observational constraint of an ionisation rate of $\zeta\lesssim 10^{-19}\mathrm{s^{-1}}$ for TW Hya \citep{cleeves-2013} since this represents a massive disk with a relatively steep radial density profile. Our results indicate that for less massive disks, which are more representative of the majority of disks around young solar-mass stars, the ionisation rate due to low energy cosmic rays should be much more significant.

Overall, we find that low energy stellar cosmic rays may be an important source of ionisation for many protoplanetary disks around young solar mass stars. An increase in the ionisation rate at the midplane of the disk may have interesting consequences for the MRI and for the launching height of magneto-centrifugally launched winds.

\section*{Acknowledgements}
This work has made use of the University of Hertfordshire's high-performance computing facility. DRL acknowledges funding from the Irish Research Council. We thank Antonella Natta for many interesting and helpful discussions. We thank the anonymous reviewer for their constructive comments.

\appendix
\label{appendix}
\section{Resolution study}
\label{appendix:a}
\begin{figure}
\centering
 \includegraphics[width=0.5\textwidth]{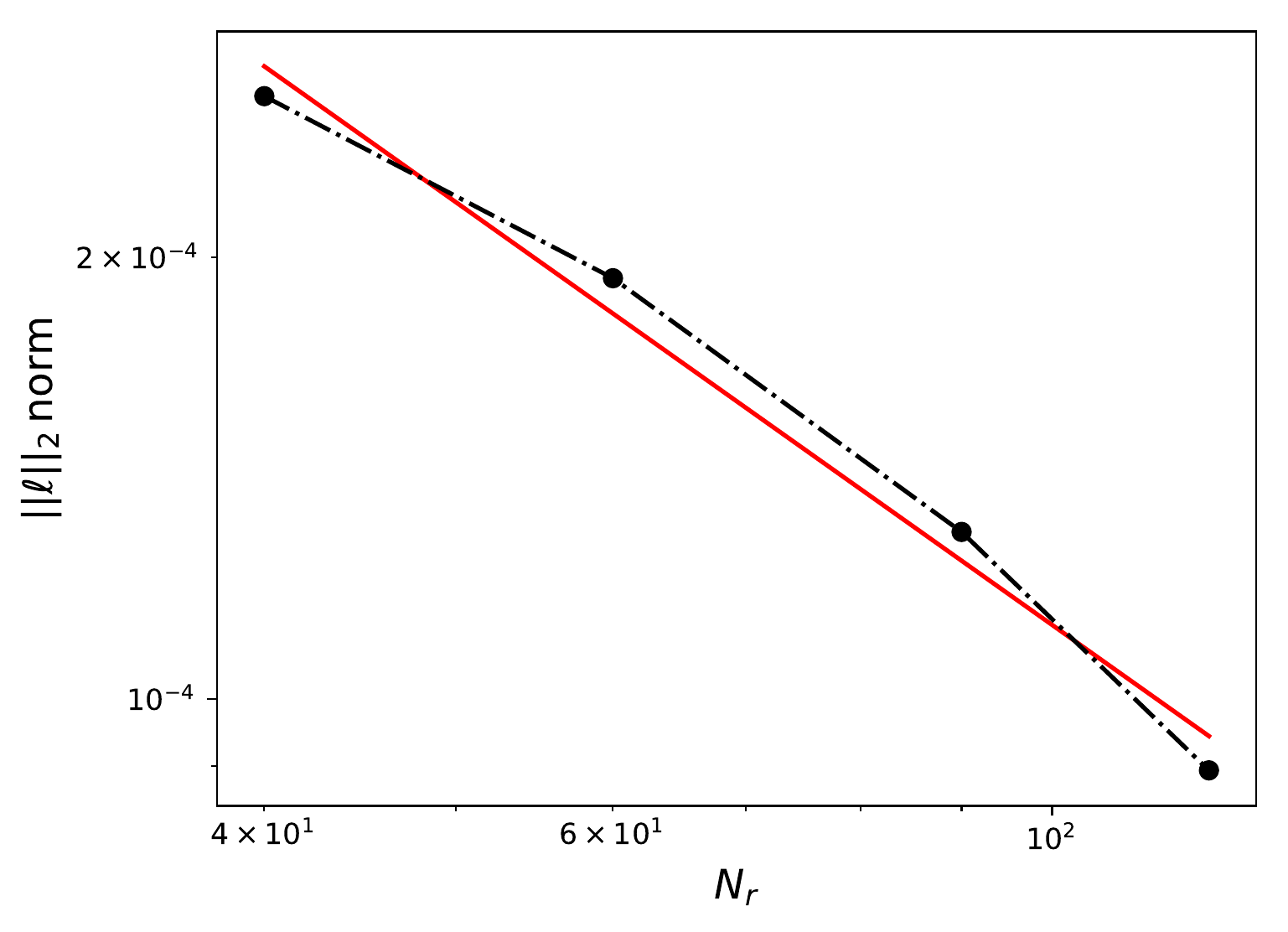}
  \caption[$||\ell||_2$ norm plotted as a function of resolution]{$||\ell||_2$ norm plotted as a function of resolution where $N_r$ is the number of grid zones in the radial direction. The number of cells in the vertical direction is also equal to $N_r$.}
\label{fig:l2-norm}
\end{figure}

We perform a resolution study using the $||\ell||_2$ norm for the standard disk parameters with $\rho_0 = 6.8\times10^{-13}\,\mathrm{g\,cm^{-3}}$ and $p=1.25$, shown in Fig.\,\ref{fig:l2-norm}. The radial and vertical extent of the disk remain the same, as given in Section\,\ref{subsubsec:rho0}. The cosmic rays are all injected at the same position and the $||\ell||_2$ norm is calculated at the same time for each of the simulations. The $||\ell||_2$ norm is defined as
\begin{equation}
||\ell(a,b)||_2 = \sqrt{\frac{1}{n}\sum^n_{i=0}|x_{i,a}-x_{i,b}|^2}
\end{equation}

\noindent where the index $i$ indicates the spatial position and the indices $a,b$ correspond to two simulations with different resolutions. Five resolutions are considered increasing the number of bins in the radial (and vertical) directions with $N_\mathrm{r}(=N_\mathrm{z})=30,40,60,90,120$. A plot of $||\ell(a,b)||_2$ on a log-log scale should yield a straight line with a slope of -1 for our scheme since, although it is second order in space, it is first order in time and the solutions have not reached a steady-state. The least-squared fitted slope of the data gives -0.95 indicating that the code is converging as expected. Furthermore the fractional difference of the cosmic ray number density between any two resolutions is less than $10^{-3}$ everywhere. We therefore conclude that our results are well resolved.

\newcommand\aj{AJ} %Astronomical Journal
\newcommand\actaa{AcA} %Acta Astronomica
\newcommand\araa{ARA\&A} %Annual Review of Astron and Astrophys
\newcommand\apj{ApJ} %Astrophysical Journal
\newcommand\apjl{ApJ} %Astrophysical Journal, Letters
\newcommand\apjs{ApJS} %Astrophysical Journal, Supplement
\newcommand\aap{A\&A} %Astronomy and Astrophysics
\newcommand\aapr{A\&A~Rev.} %Astronomy and Astrophysics Reviews
\newcommand\aaps{A\&AS} %Astronomy and Astrophysics, Supplement
\newcommand\mnras{MNRAS} %Monthly Notices of the RAS
\newcommand\pasa{PASA} %Publications of the Astron. Soc. of Australia
\newcommand\pasp{PASP} %Publications of the ASP
\newcommand\pasj{PASJ} %Publications of the ASJ
\newcommand\solphys{Sol.~Phys.} %Solar Physics
\newcommand\nat{Nature} %Nature
\newcommand\bain{Bulletin of the Astronomical Institutes of the Netherlands}
\newcommand\memsai{Mem. Societa Astronomica Italiana}
\newcommand\apss{Ap\&SS} % Astrophysics and Space Science
\newcommand\qjras{QJRAS} % Quarterly Journal of the RAS
\newcommand\pof{Physics of Fluids}
\newcommand\grl{Geophysical Research Letters}

\bibliographystyle{mn2e}
\bibliography{../../donnabib}

\label{lastpage}

\end{document}